\documentclass[twoside,12pt]{article}
\usepackage{amsmath, amsfonts, amssymb}
\usepackage{indentfirst}
\usepackage{bm}
\usepackage{epsfig}
\usepackage{graphicx}
\usepackage{epstopdf}

\begin{document}
\begin{center}
\LARGE\bf {Quasibound states of massless spin particles in Schwarzschild equivalent mediums}
\end{center}

\footnotetext{\hspace*{-.45cm}\footnotesize $^\dag$Corresponding author, E-mail:zhli@zjut.edu.cn  }

\begin{center}
\rm {Li-Qin Mi}$^{\rm a)}$, \ \ {Dandan Li}$^{\rm b)}$, \ and  \ {Zhong-Heng Li}$^{\rm c)\dagger}$
\end{center}

\begin{center}
\begin{footnotesize} \sl
${}^{\rm a)}$ {School of Physics, Zhejiang University of Technology, Hangzhou 310023, China}\\
${}^{\rm b)}$ {School of Chemical Engineering, University of Chinese Academy of Sciences, Beijing 100049, China}\\
${}^{\rm c)}${Black Hole and Gravitational Wave Group, Zhejiang University of Technology, Hangzhou 310023, China} \\
\end{footnotesize}
\end{center}



\vspace*{2mm}

\begin{center}
\begin{minipage}{15.5cm}
\parindent 20pt\footnotesize
We show that, in Schwarzschild equivalent mediums, the massless spin particles obey the same dynamical equation, from which we obtain remarkably simple formulae for the frequencies of the quasibound states. We find that the quasibound frequencies of different bosons can be identical at the same quantum number $l$, and the same is true of different fermions, but a quasibound frequency for bosons can never equal a quasibound frequency for fermions. These results mean that, in Schwarzschild equivalent mediums with the quasibound-state boundary conditions, characteristics of electromagnetic waves are the same as those for all the massless bosonic waves, thereby allowing electromagnetic waves to simulate gravitational waves. Our predictions can be tested in future experiments, building upon the successful preparation of Schwarzschild equivalent mediums.
\end{minipage}
\end{center}

\begin{center}
\begin{minipage}{15.5cm}
\begin{minipage}[t]{2.3cm}{\bf Keywords:}\end{minipage}
\begin{minipage}[t]{13.1cm}
quasibound state, theory of analogical experiment, Schwarzschild equivalent medium, gravitational wave, electromagnetic wave, massless spin particle
\end{minipage}\par\vglue8pt

\end{minipage}
\end{center}

\section{Introduction}
It is well known that quasibound states are among the most important and fundamental phenomena of physics. There has been much attention devoted to revealing characteristics of quasibound states in various black hole backgrounds in recent years. A large number of methods have been proposed. Such as the WKB method [1-3], the continued-fraction method [1,4-6], numerical method [1], the matrix method [7-10], the pseudospectral method [11,12], the asymptotic method [13-15], and the Heun-function method [16-20]. Despite extensive discussion, quasibound state problem in an equivalent medium, have not been studied to date. However, this problem is important because analogue spacetimes can provide a useful testing tool.

The idea of analogue spacetimes can be traced back almost a century ago. In 1920, Eddington [21] pointed out that the gravitational effect on light is equivalent to that produced by a suitable distribution of refractive media. In 1923, Gordon [22] studied the inverse problem, and introduced a notion of effective metric to describe the effect of a refractive index on the propagation of light. The metric was to become known as Gordon metric. The first model on analogue spcetime experiment was proposed in 1981 by Unruh [23]. Since then, a vast number of research papers on analogue gravity have been published, for instance in acoustics [24-29], electromagnetic wave guides [30], Bose-Einstein condensates [31-34], liquid helium [35,36], and graphene [37-39], etc. The developments of this field before 2011 have been introduced in the excellent review paper [40]. In recent years, especially, with the technology breakthrough of transformation optics, the photoni chips were introduced as analogical gravitational systems. From which the simulation experiments have been realized on the gravitational lens [41], Einstein ring [42], topological space of cosmic string [43], and accelerating particles in curved spacetime [44]. These studies show that the analogue gravity systems provide a new experimental platform for studying the nature of spacetime. Unlike electromagnetic waves, gravitational waves cannot be controlled with current technology. A natural question is whether researchers can mimic gravitational waves in laboratories. To our best knowledge, no experiments or theories of analogue gravitational waves was ever given for strong gravitational background.

Analogy is one of the basic thinking methods in the process of understanding objective things, its physical basis is that different physical systems obey the same dynamical evolution equations. For instance, base on the equivalence between Klein-Gordon equation for the near horizon Schwarzschild metric and the motion equation of sound waves in a convergent fluid flow, Unruh [23] established the acoustic analogue of a black hole. Therefore, similarity analyses provide cross-fertilization of ideas among different branches of science, and are the theoretical basis of analogical experiment research.

Although optical media that mimic Schwarzschild spacetime have been successfully fabricated [41], none of the currently published research results on quasibound states of Schwarzschild black holes can be experimentally verified in terrestrial laboratories using these media. This is because these studies did not use the isotropic coordinate system, which is crucial for fabricating the optical medium that mimics Schwarzschild spacetime. It is important to note that using different coordinate systems means examining the quasibound states of particles in Schwarzschild spacetime from the perspective of different observers. In this paper, we investigate quasibound states for massless scalar (spin $s=0$), Weyl neutrino (spin $s=1/2$), electromagnetic (spin $s=1$), massless Rarita-Schwinger (spin $s=3/2$), and gravitational (spin $s=2$) waves in Schwarzschild spacetime, described using the isotropic coordinate system. This spacetime can be considered an equivalent medium analogous to Schwarzschild spacetime, which we refer to as the Schwarzschild equivalent medium. We expect that our study not only solves the quasibound state problem in Schwarzschild equivalent mediums but also gives the fundamental theory for the simulation of gravitational waves in laboratories. Throughout our discussion, the particles associated the above mentioned waves are collectively called the massless spin particles for short.

This paper is organized as follows. In section 2 we study the nature of Schwarzschild equivalent mediums. In section 3 we derive the master equation governing massless spin particles in Schwarzschild equivalent mediums, and show that the angular functions of the solutions are the spin-weighted spheroidal harmonics. In section 4 we study the radial equation in Schwarzschild equivalent mediums, and discuss the form of solutions near and far the event horizon. In section 5 we obtain remarkably simple formulae for the frequencies of the quasibound States, and show that the quasibound frequencies of the massless spin particles for same statistical properties are identical at the same quantum numbers $N$ and $l$. Finally,  section 6 contains a further discussion and some concluding remarks.

\section{Schwarzschild equivalent mediums}
In isotropic coordinates, the Schwarzschild metric  is given by [45]
\begin{eqnarray}
\mathrm{d}s^{2}=\frac{(1-M/2r)^{2}}{(1+M/2r)^{2}}\mathrm{d}t^{2}
-(1+\frac{M}{2r})^{4}(\mathrm{d}r^{2}+r^{2}\mathrm{d}\theta^{2}+r^{2}\sin^{2}\theta \mathrm{d}\varphi^{2}),
\label{eq:1}
\end{eqnarray}
where $M$ is the mass of the black hole. The event horizon in isotropic coordinates is located at $r_{H}=M/2$. It is well-known that, for the Schwarzschild metric in the standard form, the most obvious characteristic at the event horizon is the reversal there of the roles of $t$ and $r$ as timelike and spacelike coordinates. But when the isotropic coordinates are adopted, this argument fails. Metric  (1) shows that the $t$ direction is timelike and the $r$ direction is spacelike,  both in the region $r>M/2$ and in the region $r<M/2$. Actually, the metric (1) does not cover the interior region of the Schwarzschild black hole; it covers the exterior region twice [46].

In 1923, Gordon [22] introduced the metric describing an equivalent medium, which is
\begin{eqnarray}
g_{\mu\nu}=\eta_{\mu\nu}+(\frac{1}{n^{2}}-1)u_{\mu}u_{\nu},
\label{eq:2}
\end{eqnarray}
where $\eta_{\mu\nu}$ is the flat metric, $n$ is the refractive index, and $u^{\mu}$ is the 4-velocity of the medium. This is the famous Gordon metric. In original Gordon metric, $n$ and $u_{\mu}$ are constants. To simulate a real spacetime, the following generalization of the Gordon is widely used [47]:
\begin{eqnarray}
g_{\mu\nu}=\Omega^{2}[\eta_{\mu\nu}+(\frac{1}{n^{2}}-1)u_{\mu}u_{\nu}].
\label{eq:3}
\end{eqnarray}
Here $\Omega$ is called the conformal factor, and the quantities $\Omega$, $n$, and $u_{\mu}$ are allowed to be space and time dependent.

In the rest frame of the medium $u^{\mu}=(1, 0, 0, 0)$, Eq. (3) reads
\begin{eqnarray}
g_{\mu\nu}=\Omega^{2}[\eta_{\mu\nu}+(\frac{1}{n^{2}}-1)\delta_{\mu}^{0}\delta_{\nu}^{0}].
\label{eq:4}
\end{eqnarray}
In spherical polar coordinates $r$, $\theta$, $\varphi$, the line element of a conformal Gordon spacetime can be written as
\begin{eqnarray}
\mathrm{d}s^{2}=\Omega^{2}(r)[\frac{1}{n^{2}(r)}\mathrm{d}t^{2}-(\mathrm{d}r^{2}+r^{2}\mathrm{d}\theta^{2}+r^{2}\sin^{2}\theta \mathrm{d}\varphi^{2})].\nonumber\\
\label{eq:5}
\end{eqnarray}

The comparison of Eq. (5) with Eq. (1) shows that, for the Schwarzschild metric, the conformal factor and the index of refraction take the forms
\begin{equation}
\Omega=(1+\frac{M}{2r})^{2},
\label{eq:6}
\end{equation}
and
\begin{equation}
n=(1+\frac{M}{2r})^{3}(1-\frac{M}{2r})^{-1}.
\label{eq:7}
\end{equation}

Metric (5) with Eqs. (6) and (7) is also called the Schwarzschild equivalent medium. This metric is important because it has been realized in  an optical medium, which portray a good analogical gravitational system. This means that if a spin particle has Maxwell electromagnetic wavelike properties in Schwarzschild spacetime, then propagation of electromagnetic waves in the analogical gravitational system can be used to simulate the properties this spin particle. This is a main motivations for our work.

Starting from the calculation of the spin coefficients [see Eqs. (A1) and (A3)] we can prove that a Schwarzschild equivalent medium is of Petrov type D by Goldberg-Sachs theorem [48]. The values of the non-vanishing Weyl scalar $\psi_{2}$ and the Ricci scalar $R$ [for definition see Eqs. (A4) and (A5)] can be given by
\begin{eqnarray}
\psi_{2}=\frac{1}{6\Omega^{2}}[-\frac{n''}{n}+2(\frac{n'}{n})^{2}+\frac{1}{r}\frac{n'}{n}],
\label{eq:8}
\end{eqnarray}
\begin{eqnarray}
&&R=\frac{2}{\Omega^{2}}[3(-\frac{\Omega''}{\Omega}+\frac{\Omega'}{\Omega}\frac{n'}{n}-\frac{2}{r}\frac{\Omega'}{\Omega})
+\frac{n''}{n}-2(\frac{n'}{n})^{2}+\frac{2}{r}\frac{n'}{n}],\nonumber\\
\label{eq:9}
\end{eqnarray}
where the prime denots the derivative with respect to $r$.

\section{Unified wave equation for massless spin particles}

Now we will consider the perturbation equation for massless fields of various spin $s\leq2$ on the Schwarzschild equivalent medium background. In general, the field equations of spin 1/2 1, 3/2, and 2 are not accurately separable, but in all the type-D metrics, the massless field equations of spin 1/2, 1, 3/2, and 2 can be decoupled in the case of perturbations [49-52]. The findings indicate that each spin state of a particle corresponds to a equation, resulting in a total of eight equations, as detailed in Appendix B. To enhance the identification of similarities among their solutions, the solutions must be articulated in a unified form. This necessitates the integration of these equations into a single statement (master equation).

In Appendix B, each pair of Eqs. (B3)-(B6) represents the Weyl neutrino, electromagnetic, massless Rarita-Schwinger, and gravitational perturbations, respectively. In each pair, the first equation corresponds to the spin state $p=s$, while the second corresponds to $p=-s$. A careful examination of (B3)-(B6) readily allows the equations describing the spin states $p=s$ to be unified into the following form:
\begin{eqnarray}\label{10}
&\{[D-(2s-1)\varepsilon+\bar{\varepsilon}-2s\rho-\bar{\rho}](\Delta-2s\gamma+\mu) \nonumber\\
&-[\delta+\bar{\pi}-\bar{\alpha}-(2s-1)\beta-2s\tau](\bar{\delta}+\pi-2s\alpha) \nonumber\\
&-(2s-1)(s-1)\psi_{2}\}\Phi_{s}=0,
\end{eqnarray}
and those for the spin states $ p = -s $ into the following form:
\begin{eqnarray}\label{11}
&\{[\Delta+(2s-1)\gamma-\bar{\gamma}+2s\mu+\bar{\mu}](D+2s\varepsilon-\rho) \nonumber\\
&-[\bar{\delta}-\bar{\tau}+\bar{\beta}+(2s-1)\alpha+2s\pi](\delta-\tau+2s\beta) \nonumber\\
&-(2s-1)(s-1)\psi_{2}\}\Phi_{-s}=0.
\end{eqnarray}
Here, $\Phi_{s}$ and $\Phi_{-s}$ denote the wave functions corresponding to the particle's spin states $p=s$ and $p=-s$, respectively. The Greek letters such as $\rho$ represent the spin coefficients, whose definitions are provided in Eq. (A1) of Appendix A. $D, \Delta,$ and $\delta$ are the directional derivatives defined by the equations
\begin{equation}\label{12}
D=l^{\mu}\partial_{\mu}, \quad \Delta=n^{\mu}\partial_{\mu}, \quad \delta=m^{\mu}\partial_{\mu}, \quad \bar{\delta}=\bar{m}^{\mu}\partial_{\mu};
\end{equation}
where $l^{\mu}$, $n^{\mu}$, $m^{\mu}$ and $\bar{m}^{\mu}$ are the null tetrads. For Schwarzschild equivalent mediums, the null tetrads and spin coefficients are given by Eqs. (A2) and (A3). Substituting Eqs. (12), (A2) and (A3) into Eqs. (10) and (11), and applying the transformation
\begin{equation}
\Phi_{p}=(\Omega r)^{(p-s)}\Psi_{p}.
\label{eq:13}
\end{equation}
we can combine Eqs. (10) and (11) into a single master equation, which can be written in the elegant form
\begin{equation}
[(\nabla^{\mu}+pL^{\mu})(\nabla_{\mu}+pL_{\mu})-4p^{2}\psi_{2}+\frac{1}{6}R]\Psi_{p}=0,
\label{eq:14}
\end{equation}
where
\begin{eqnarray}
L^{t}&=&\frac{n}{\Omega^{2}}(\frac{n'}{n}+\frac{1}{r}),\nonumber\\
L^{r}&=&\frac{1}{\Omega^{2}}(2\frac{\Omega'}{\Omega}-\frac{n'}{n}-\frac{1}{r}),\nonumber\\
L^{\theta}&=&0,\nonumber\\
L^{\varphi}&=&-\frac{1}{\Omega^{2}r^{2}}\frac{\mathrm{i} \cos\theta}{\sin^{2}\theta}.
\label{eq:15}
\end{eqnarray}
Here, $\nabla_{\mu}$ denotes the covariant derivative in the metric $g_{\mu\nu}$.

Equation (14) tells us that the wave functions of different massless spin particles $\Psi_{p}$ obey the same dynamical equation (also called wave equation).  In view of the completely different nature of these particles, it is surprising (and delightful) that their dynamical equations have such a similar structure in Schwarzschild equivalent mediums. Evidently, when $p=0$, Eq. (14) is just the (conformally invariant) massless scalar field equation. Therefore, Eq. (14) governs not only the massless fields of spin 1/2, 1, 3/2, and 2, but also the scalar field. Equation (14) is the fundamental result, from which we can study the similarity of the wave functions for different massless spin particles in the Schwarzschild equivalent medium.

The fact that the exact solution of Eq. (14) has the form
\begin{equation}
\Psi_{p}=\mathrm{e}^{-\mathrm{i}\omega t}S_{p}(\theta,\varphi)R_{p}(r).
\label{eq:16}
\end{equation}

The equation for $S_{p}$ is
\begin{eqnarray}
\big[\frac{1}{\sin\theta}\frac{\partial}{\partial\theta}\big(\sin\theta\frac{\partial}{\partial\theta}\big)&+&\frac{1}{\sin^{2}\theta}\frac{\partial^{2}}{\partial\varphi^{2}}
+\frac{2\mathrm{i}p\cos\theta}{\sin^{2}\theta}\frac{\partial}{\partial\varphi}\nonumber\\
&-&p^{2}\cot^{2}\theta+p+(l-p)(l+p+1)\big]S_{p}(\theta,\varphi)=0.
\label{eq:17}
\end{eqnarray}
Then the angular part of the wave function is a spin-weighted spherical harmonic [53,54], namely
\begin{eqnarray}
S_{p}(\theta, \varphi)&\equiv &\,_{p}Y_{lm}(\theta,\varphi)
=\big[\frac{(2l+1)}{4\pi}\frac{(l+m)!(l-m)!}{(l+p)!(l-p)!}\big]^{1/2}\mathrm{e}^{\mathrm{i}m\varphi}\big(\sin\frac{\theta}{2}\big)^{2l}\nonumber\\
&\cdot &\sum_{k}\left(\begin{array}{lll}
l-p\\
\,\,\,\,k
\end{array}
\right)\left(\begin{array}{lll}
\,\,\,\,\,\,l+p\\
k+p-m
\end{array}
\right)(-1)^{l-k-p}\big(\cot\frac{\theta}{2}\big)^{2k+p-m},
\label{eq:18}
\end{eqnarray}
where $l$, and $m$ are integers satisfying the inequalities $l\geq s$, $-l\leq m\leq l$. The spin-weighted spherical harmonics satisfy the orthonormality
and completeness relations
\begin{equation}
\int_{0}^{2\pi}\mathrm{d}\varphi\int_{0}^{\pi}\mathrm{d}\theta\,_{p}Y_{lm}(\theta,\varphi)\,_{p}\bar{Y}_{l'm'}(\theta,\varphi)\sin\theta=\delta_{ll'}\delta_{mm'},
\label{eq:19}
\end{equation}
and
\begin{equation}
\sum_{m=-l}^{l}\mid\,_{p}Y_{lm}(\theta,\varphi)\mid^{2}=\frac{2l+1}{4\pi},
\label{eq:20}
\end{equation}
respectively.

It is obvious that the actual shape of the metric functions affects only the radial function $R_{p}(r)$, which has the form
\begin{equation}
R_{p}(r)=A\,\Omega^{2p-1}n^{(1-2p)/2}r^{-(p+1)}g(r),
\label{eq:21}
\end{equation}
where $A$ is some constant, the functions $g(r)$ has to satisfy
\begin{eqnarray}
\frac{\mathrm{d}^{2}g(r)}{\mathrm{d}r^{2}}&+&\big[\omega^{2}n^{2}+2\mathrm{i}\omega p n(\frac{n'}{n}+\frac{1}{r})-\frac{1}{3}(2p-1)(2p+1)\nonumber\\
&\cdot &\big(\frac{1}{2}\frac{n''}{n}-\frac{1}{4}(\frac{n'}{n})^{2}+\frac{1}{r}\frac{n'}{n}\big)-\frac{l(l+1)}{r^{2}}\big]g(r)=0.
\label{eq:22}
\end{eqnarray}
This equation shows that the functions $g(r)$ is not affected by the conformal factor $\Omega(r)$.

Substituting Eq. (7) into Eq. (22), we obtain
\begin{eqnarray}
\frac{\mathrm{d}^{2}g(r)}{\mathrm{d}r^{2}}&+&\big\{\omega^{2}\frac{(1+M/2r)^{6}}{(1-M/2r)^{2}}-\mathrm{i}\omega p\big[\frac{2M(M+2r)^{2}}{r^{2}(M-2r)^{2}}-\frac{(M+2r)^{2}}{2r^{3}}\big]\nonumber\\
&-&(4p^{2}-1)\frac{4M^{2}}{(M^{2}-4r^{2})^{2}}-\frac{l(l+1)}{r^{2}}\big\}g(r)=0.
\label{eq:23}
\end{eqnarray}
An exact solution of this equation would be very difficult to obtain. However, as we shall see, the most important features of the solution can be found by the uniqueness theorem.

\section{Radial functions near and far the event horizon}

It is clear that the coefficient in Eq. (23) are single-valued analytic functions in the interval ($M/2$, $\infty$). The uniqueness theorem tells us that there is only one analytic function satisfying the differential equation and the given boundary conditions. The clear inference is that the asymptotic behavior of the solutions of the equation is therefore dominated by the solutions of the asymptotic equation. For this reason, below we study the asymptotic equation of Eq. (23) and its solutions.

\subsection{Solutions in the near zone}

In the vicinity of the event horizon, Eq. (23) should have the form
\begin{equation}
\frac{\mathrm{d}^{2}g}{\mathrm{d}r^{2}}+\big[\omega^{2}+\frac{a_{1}}{r-M/2}+\frac{a_{2}}{(r-M/2)^{2}}\big]g=0,
\label{eq:24}
\end{equation}
where
\begin{equation}
a_{1}=-\frac{2a_{2}}{M}, \quad a_{2}=\frac{1}{4}-(p+\mathrm{i}4M\omega)^{2}.
\label{eq:25}
\end{equation}

It should be noted that if we restrict our consideration to the region near the event horizon, we can safely neglect the $\omega^{2}$ term inside the square brackets in Eq. (24). This is because whether or not the $\omega^{2}$ term is present, the form of the solution to Eq. (24) near the outer event horizon remains unchanged-specifically, all solutions reduce to Eq. (48) in Section 5. In other words, retaining the $\omega^{2}$ term does not affect the solution's behavior near the event horizon.

Note that as $r\rightarrow\infty$, the limiting form of Eq. (24) becomes $\frac{\mathrm{d}^{2}g}{\mathrm{d}r^{2}}+\omega^{2}g=0$. This is identical to the equation derived from the Eq. (23) under the same limit. The subtlety in retaining the $\omega^{2}$ term stems from the fact that Eq. (24) serves not only as an approximation of Eq. (23) near the event horizon, but also as its asymptotic form when $r\rightarrow\infty$.

If we introduce a new variable
\begin{equation}
x=r-\frac{M}{2},
\label{eq:26}
\end{equation}
in terms of $x$, the radial equation (24) in the vicinity of the horizon takes the form
\begin{equation}
\frac{\mathrm{d}^{2}g}{\mathrm{d}x^{2}}+\big[\omega^{2}+\frac{a_{1}}{x}+\frac{a_{2}}{x^{2}}\big]g=0.
\label{eq:27}
\end{equation}

Again, we transform to new variable and function. First we put
\begin{equation}
z=\mathrm{i}2\omega x,
\label{eq:28}
\end{equation}
and then we define
\begin{equation}
g(r)=\mathrm{e}^{-\mathrm{i}\omega x}x^{\frac{1}{2}(1+\sqrt{1-4a_{2}})}u.
\label{eq:29}
\end{equation}
Thus Eq. (27) transforms to
\begin{equation}
z\frac{\mathrm{d}^{2}u}{\mathrm{d}z^{2}}+(1+\sqrt{1-4a_{2}}-z)\frac{\mathrm{d}u}{\mathrm{d}z}-[\frac{1}{2}(1+\sqrt{1-4a_{2}})+\frac{\mathrm{i}a_{1}}{2\omega}]u=0.
\label{eq:30}
\end{equation}
Clearly, by setting
\begin{equation}
\gamma=1+\sqrt{1-4a_{2}}=1+2(\mathrm{i}4M\omega+p),
\label{eq:31}
\end{equation}
\begin{eqnarray}
\alpha &=&\frac{1}{2}(1+\sqrt{1-4a_{2}})+\frac{\mathrm{i}a_{1}}{2\omega}=\frac{1}{2}(1+\sqrt{1-4a_{2}})-\frac{\mathrm{i}a_{2}}{M\omega}\nonumber\\
&=&\frac{1}{2}[1+2(\mathrm{i}4M\omega+p)]+\frac{\mathrm{i}}{M\omega}[(\mathrm{i}4M\omega+p)^{2}-\frac{1}{4}],
\label{eq:32}
\end{eqnarray}
Eq. (30) becomes the standard degenerate hypergeometric equation:
\begin{equation}
z\frac{\mathrm{d}^{2}u}{\mathrm{d}z^{2}}+(\gamma-z)\frac{\mathrm{d}u}{\mathrm{d}z}-\alpha u=0,
\label{eq:33}
\end{equation}
As is well known, if $\gamma$ is not an integer, then the general solution of the degenerate hypergeometric equation has the form
\begin{eqnarray}
u=C_{1}F(\alpha, \gamma; z)+C_{2}z^{1-\gamma}F(\alpha+1-\gamma, 2-\gamma; z),
\label{eq:34}
\end{eqnarray}
where
\begin{equation}
F(\alpha, \gamma; z)=\sum^{\infty}_{k=1}\frac{(\alpha)_{k}}{k!(\gamma)_{k}}z^{k}.
\label{eq:35}
\end{equation}
Here $F$ is called the degenerate hypergeometric function or the Kummer function.

Therefore, if $1+\sqrt{1-4a_{2}}$ is not an integer, then the general solution of Eq. (27) has the form
\begin{eqnarray}
g&=&\mathrm{e}^{-\mathrm{i}\omega x}x^{\frac{1}{2}(1+\sqrt{1-4a_{2}})}[C_{1}F(\frac{1}{2}(1+\sqrt{1-4a_{2}})-\frac{\mathrm{i}a_{2}}{M\omega}, 1+\sqrt{1-4a_{2}}; \mathrm{i}2\omega x)\nonumber\\
&+&C_{2}x^{-\sqrt{1-4a_{2}}}F(\frac{1}{2}(1-\sqrt{1-4a_{2}})-\frac{\mathrm{i}a_{2}}{M\omega}, 1-\sqrt{1-4a_{2}}; \mathrm{i}2\omega x)],
\label{eq:36}
\end{eqnarray}
where $C_{1}$ and $C_{2}$ are arbitrary constants.

\subsection{Solutions in the far zone}

Our this step is to find the behaviour of $g(r)$ at large $r$. In this case, the differential equation (23) becomes
\begin{equation}
\frac{\mathrm{d}^{2}g}{\mathrm{d}r^{2}}+(\omega^{2}+\frac{b_{1}}{r}+\frac{b_{2}}{r^{2}})g=0,
\label{eq:37}
\end{equation}
where
\begin{equation}
b_{1}=2\omega(2\omega M+\mathrm{i}p), \quad b_{2}=\frac{15}{2}\omega^{2}M^{2}-l(l+1).
\label{eq:38}
\end{equation}

Using a method similar to that of Eq. (27), we transform equation (37) through the following changes:
\begin{equation}
z=\mathrm{i}2\omega r,
\label{eq:39}
\end{equation}
\begin{equation}
g(r)=\mathrm{e}^{-\mathrm{i}\omega r}r^{\frac{1}{2}(1+\sqrt{1-4b_{2}})}u.
\label{eq:40}
\end{equation}
This converts Eq. (37) into
\begin{equation}
z\frac{\mathrm{d}^{2}u}{\mathrm{d}z^{2}}+(1+\sqrt{1-4b_{2}}-z)\frac{\mathrm{d}u}{\mathrm{d}z}-[\frac{1}{2}(1+\sqrt{1-4b_{2}})+\frac{\mathrm{i}b_{1}}{2\omega}]u=0.
\label{eq:41}
\end{equation}

If we set $\gamma=1+\sqrt{1-4b_{2}}$ and $\alpha=(1+\sqrt{1-4b_{2}})/2+\mathrm{i}b_{1}/(2\omega)$, Eq. (41) becomes the standard degenerate hypergeometric equation (33) as well. A natural conclusion is that, if $1+\sqrt{1-4b_{2}}$ is not an integer, the general solution of Eq. (37) can be expressed in the form
\begin{eqnarray}
g&=&\mathrm{e}^{-\mathrm{i}\omega r}r^{(1+\sqrt{1-4b_{2}})/2}
\big[D_{1}F(-p+\mathrm{i}2\omega M+\frac{1}{2}(1+\sqrt{1-4b_{2}}), 1+\sqrt{1-4b_{2}}; \mathrm{i}2\omega r)\nonumber\\
&+&D_{2}r^{-\sqrt{1-4b_{2}}}
F(-p+\mathrm{i}2\omega M+\frac{1}{2}(1-\sqrt{1-4b_{2}}), 1-\sqrt{1-4b_{2}}; \mathrm{i}2\omega r)\big],
\label{eq:42}
\end{eqnarray}
where $D_{1}$ and $D_{2}$ are arbitrary constants.

The existence and uniqueness of the solution of differential equation ensure that the behavior of the solutions of Eq. (23) near the event horizon and far from the event horizon can be described by Eqs. (36) and (42), respectively.

When solutions to a differential equation are not constrained by boundary conditions, the independent particular solutions of Eqs. (24) and (37) exhibit a common characteristic as $r\rightarrow\infty$: one particular solution contains the factor $\mathrm{e}^{\mathrm{i}\omega r}$, while the other contains $\mathrm{e}^{-\mathrm{i}\omega r}$. The particular solutions of Eqs. (24) and (37) sharing the same exponential factor represent approximations of one independent analytic particular solutions of Eq. (23) in the near-region and far-region respectively.  In other words, as $r\rightarrow\infty$, the two particular solutions in Eqs. (30) and (34) containing the factor $\mathrm{e}^{\mathrm{i}\omega r}$ approximate a single independent particular solution of Eq. (23) -- one localized near the event horizon and the other in the far region. Similarly, the two particular solutions in Eqs. (36) and (42) with $\mathrm{e}^{-\mathrm{i}\omega r}$ approximate another independent particular solution of Eq. (23), with analogous spatial distributions. Thus, retaining the $\omega^{2}$ term provides an effective method for matching near-region and far-region solutions.

\section{Quasibound frequencies}

The uniqueness theorem for second-order differential equations states that if all coefficients are single-valued and analytic within a region, then given two specified boundary conditions, there exists a unique single-valued analytic solution in that domain. The theoretical existence of this solution is guaranteed regardless of whether mathematical techniques are used to explicitly find it. Therefore, the spectrum is determined solely by the boundary conditions. Any method that artificially imposes additional constraints at non-boundary locations to determine the spectrum would contradict the differential equation's uniqueness theorem.

The primary focus is on ensuring consistency between the solutions in the near-horizon region, far-horizon region, and the exact analytic solution across the entire domain, rather than on the consistency of these solutions in other regions. This is because, in those regions, the near-horizon and far-horizon solutions differ significantly from the exact analytic solution. Consequently, discussing the properties of these solutions in other regions holds no theoretical or practical value.

Specifically, requiring the near-region solution (36) and far-region solution (42) to coincide at an intermediate point in the overlapping region effectively imposes an extra constraint. This approach may initially appear reasonable for determining quasibound frequencies, but it violates the uniqueness theorem for second-order differential equations. Any method that artificially introduces additional conditions within the domain (excluding boundaries) to determine the quasibound frequencies lack rigorous justification; the resulting wave function may fail to satisfy both boundary conditions simultaneously. This occurs because a general solution to a second-order differential equation contains only two arbitrary constants, making it fundamentally impossible for a single solution to satisfy three independent conditions.

As mentioned baove, Eqs. (36) and (42) are consistent with the behavior of the solutions of Eq. (23) at small $r-M/2$ and larger $r$, respectively. The asymptotic behavior of the solutions of Eq. (23) is therefore dominated by Eqs. (36) and (42). Thus we can study the quasibound states and frequencies by means of Eqs. (36) and (42).

It is widely known that the quasinormal-mode boundary conditions are that the wave is purely ingoing on the event horizon and purely outgoing at spatial infinity, while quasibound-state boundary conditions are that the wave is purely ingoing on the event horizon and tend to zero at spatial infinity, that is to say that the complete radial function, $(\Omega r)^{p-s}R_{p}$, must satisfy the following boundary conditions [16-20]:
\begin{equation}
(\Omega r)^{p-s}R_{p}\sim\
\left\{
\begin{array}{ c c }
\mathrm{e}^{-\mathrm{i}\omega r_{\ast}}, & r_{\ast}\rightarrow -\infty, \\
0, & r_{\ast}\rightarrow \infty, \\
\end{array}
\right.
\label{eq:43}
\end{equation}
where $r_{*}$ is called the tortoise coordinate. It is determined by the equation:
\begin{equation}
\partial^{\mu}v \partial_{\mu}v=0, \quad \partial^{\mu}u \partial_{\mu}u=0.
\label{eq:44}
\end{equation}
Here $v$ and $u$ are the Eddington-Finkelstein null coordinates, which take the form
\begin{equation}
v=t+r_{*}, u=t-r_{*}.
\label{eq:45}
\end{equation}

By substituting Eq. (45) into Eq. (44) and using metric (1), we derive the exact form of the tortoise coordinate for Schwarzschild equivalent mediums:
\begin{equation}
r_{*}=r+\frac{r_{H}^{2}}{r}-\frac{1}{2\kappa}\ln\frac{r}{r_{H}}+\frac{1}{\kappa}\ln|\frac{r-r_{H}}{r_{H}}|.
\label{eq:46}
\end{equation}
Note that $\kappa=1/(4M)$ is the surface gravity (not to be confused with the spin coefficient). When $r\rightarrow r_{H}$, we have
\begin{equation}
r_{*}\approx\frac{1}{\kappa}\ln|\frac{r-r_{H}}{r_{H}}|.
\label{eq:47}
\end{equation}

Near the event horizon $r_{*}\rightarrow -\infty$, $r\rightarrow M/2$, the complete radial function has the following asymptotic behavior at the exterior event horizon,
\begin{eqnarray}
(\Omega r)^{p-s}R_{p}&=&C_{1}(r-\frac{M}{2})^{2p+\mathrm{i}4M\omega}+C_{2}(r-\frac{M}{2})^{-\mathrm{i}4M\omega}\nonumber\\
&=&C_{1}\mathrm{e}^{(2p\kappa+\mathrm{i}\omega)r_{*}}+C_{2}\mathrm{e}^{-\mathrm{i}\omega r_{*}},
\label{eq:48}
\end{eqnarray}
where all the remaining constants are  included in $C_{1}$ and $C_{2}$. The boundary condition at $r=M/2$ requires that $C_{1}=0$, and hence we have
\begin{equation}
g=C_{2}x^{1-\frac{\gamma}{2}}F(\alpha+1-\gamma, 2-\gamma; \mathrm{i}2\omega x)  \quad (\textmd{for small $r-\frac{M}{2}$}).
\label{eq:49}
\end{equation}

Once we have determined the solution near the event horizon, a natural question arises: which of the two particular solutions in Eq. (42) corresponds to the given particular solution in Eq. (36)? To answer  this question, we retain the first term in square brackets in Eq. (24). When $r\rightarrow\infty$, the limiting function of Eq. (49) contains the factor $\mathrm{e}^{\mathrm{i}\omega r}$, and similarly, the limiting function of the second particular solution in Eq. (42) also contains this factor as $r\rightarrow\infty$. Therefore, we conclude that the second particular solution in Eq. (42) and Eq. (49) correspond to the particular solution of Eq. (23) that contains the factor $\mathrm{e}^{\mathrm{i}\omega r}$ in the asymptotic limit of $r\rightarrow\infty$.

To match Eq. (49), we should have that the value of $D_{1}$ in Eq. (42) is zero, then one finds that for large values of $r$,
\begin{eqnarray}
g=D_{2}r^{(1-\sqrt{1-4b_{2}})/2}\mathrm{e}^{-\mathrm{i}\omega r}
F(1-p+\mathrm{i}2\omega M-\frac{1+\sqrt{1-4b_{2}}}{2}, 1-\sqrt{1-4b_{2}}; \mathrm{i}2\omega r)\big].
\label{eq:50}
\end{eqnarray}

To obtain expressions for the quasibound frequencies we need the $r\rightarrow\infty$ (or $r_{*}\rightarrow\infty$) asymptotic form of the complete radial function $(\Omega r)^{p-s}R_{p}(r)$ which is given as
\begin{equation}
(\Omega r)^{p-s}R_{p}(r)\sim D_{2}r^{-(s+p+1)}\mathrm{e}^{\mathrm{i}\omega r}.
\label{eq:51}
\end{equation}

It should be noted that Eq. (51) has the form $\mathrm{e}^{\mathrm{i}\omega r}/r$ when $p=-s$, which is an outgoing spherical wave we don't want. Because the physical requirement is that the imaginary parts of the quasibound frequencies are negative (due to the time factor $\mathrm{e}^{-\mathrm{i}\omega t}$ being finite as $t\rightarrow\infty$), this leads to an infinite value of Eq. (51) in the limit $r\rightarrow\infty$, and thus does not satisfy the boundary condition (43) at spatial infinity. On the other hand, the Kummer function in Eq. (50) has the asymptotic form $F\sim r^{-p+(1-\sqrt{1-4b_{2}})/2}\mathrm{e}^{\mathrm{i}2\omega r}$ for very large $r$. The Kummer function's contribution to Eq. (51) evidently includes the factor $\mathrm{e}^{\mathrm{i}2\omega r}$. Therefore, there is only one way to wiggle out of this: for quasibound states, the Kummer function must be forcibly truncated and become Laguerre polynomials. Thus, asymptotic form of the complete radial function is the product of a polynomial and $\mathrm{e}^{-\mathrm{i}\omega r}$ [this can also be seen directly from Eq. (50)], which becomes zero as $r\rightarrow\infty$, thereby satisfying the boundary condition.

The condition for the Kummer's series in Eq. (50) to become the Laguerre polynomials is

\begin{equation}
1-\frac{1}{2}(1+\sqrt{1-4b_{2}})-p+\mathrm{i}2M\omega=-k,
\label{eq:52}
\end{equation}
where $k=0, 1, 2, 3...$. This gives remarkably simple formulae for the frequencies of the quasibound states (quasibound frequencies):

\begin{equation}
\frac{7}{2}M\omega=
\left\{
\begin{array}{ c c }
-\mathrm{i}N+\frac{1}{4}\sqrt{14(2l+1)^{2}-30N^{2}}, & \textmd{if}\ 2l+1>\sqrt{15/7}N;\\
-\mathrm{i}[N\pm\frac{1}{4}\sqrt{-14(2l+1)^{2}+30N^{2}}], & \textmd{if}\ N< 2l+1<\sqrt{15/7}N;\\
-\mathrm{i}[N+\frac{1}{4}\sqrt{-14(2l+1)^{2}+30N^{2}}], & \textmd{if}\ 2l+1\leq N.
\end{array}
\right.
\label{eq:53}
\end{equation}
where $N=2k-2p+1=1, 2, 3,...$. Note that other solutions of Eq. (52) are unphysical. Obviously all quasibound frequencies result in damped modes. Degrees of damping depend on the values of $l$ and $N$. Underdamped cases occur only for $2l+1>\sqrt{15/7}N$, which exhibit oscillatory behavior and the damping absent when $N=0$. However, the case $N=0$ must be excluded because of the boundary condition at $r\rightarrow\infty$. As shown in Fig. 1, a finite number of underdamped modes can arise for a given value of $l$.

\begin{figure}
\centering
\includegraphics[width=.4\textwidth,trim= 0 0 0 0,clip]{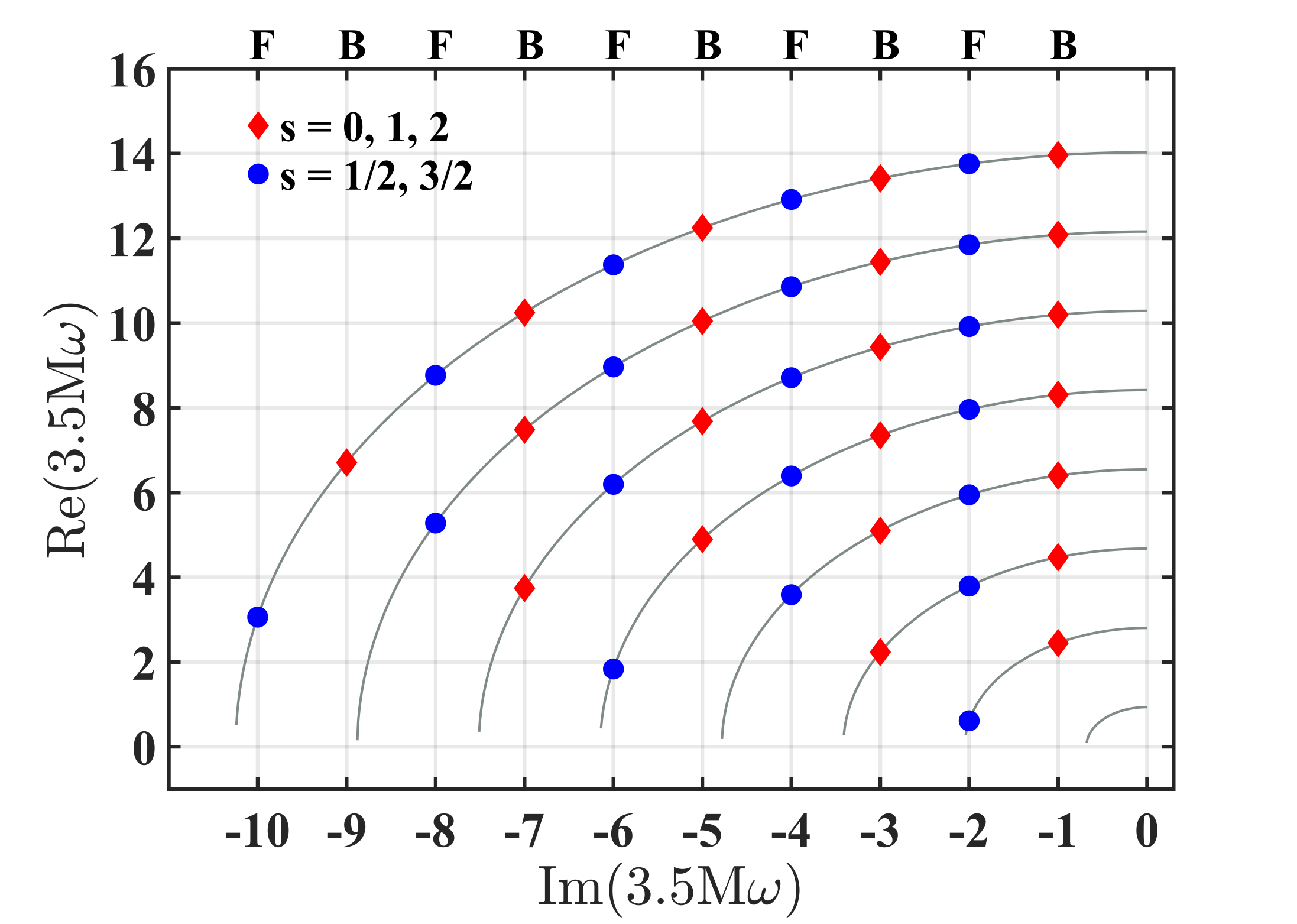}
\caption{\label{fig:1} Quasibound frequencies in units of $1/(3.5M)$ for $2l+1>\sqrt{15/7}N$. The absolute value of $Im(3.5M\omega)$ is an odd integer for bosons (B) and an even integer for fermions (F). All frequencies on the same curve have the same quantum number $l$. The curves towards to the top left of the graph correspond to larger $l$ valus, which are $0, 1, 2, 3, 4, 5, 6$, and $7$.}
\end{figure}


When $2l+1<\sqrt{15/7}N$, $\omega$ takes only discrete imaginary values, which leads to an exponentially decreasing function of distance (also time). The behavior is called the overdamped modes. Of especial interest is that there are two branches of overdamped modes for $N< 2l+1<\sqrt{15/7}N$. As similar to the underdamped modes, for a given value of $l$, only a finite number of overdamped modes with the smaller $\omega$ of two branches can arise as shown in Fig. 2.

\begin{figure}
\centering
\includegraphics[width=.4\textwidth,trim= 0 0 0 0,clip]{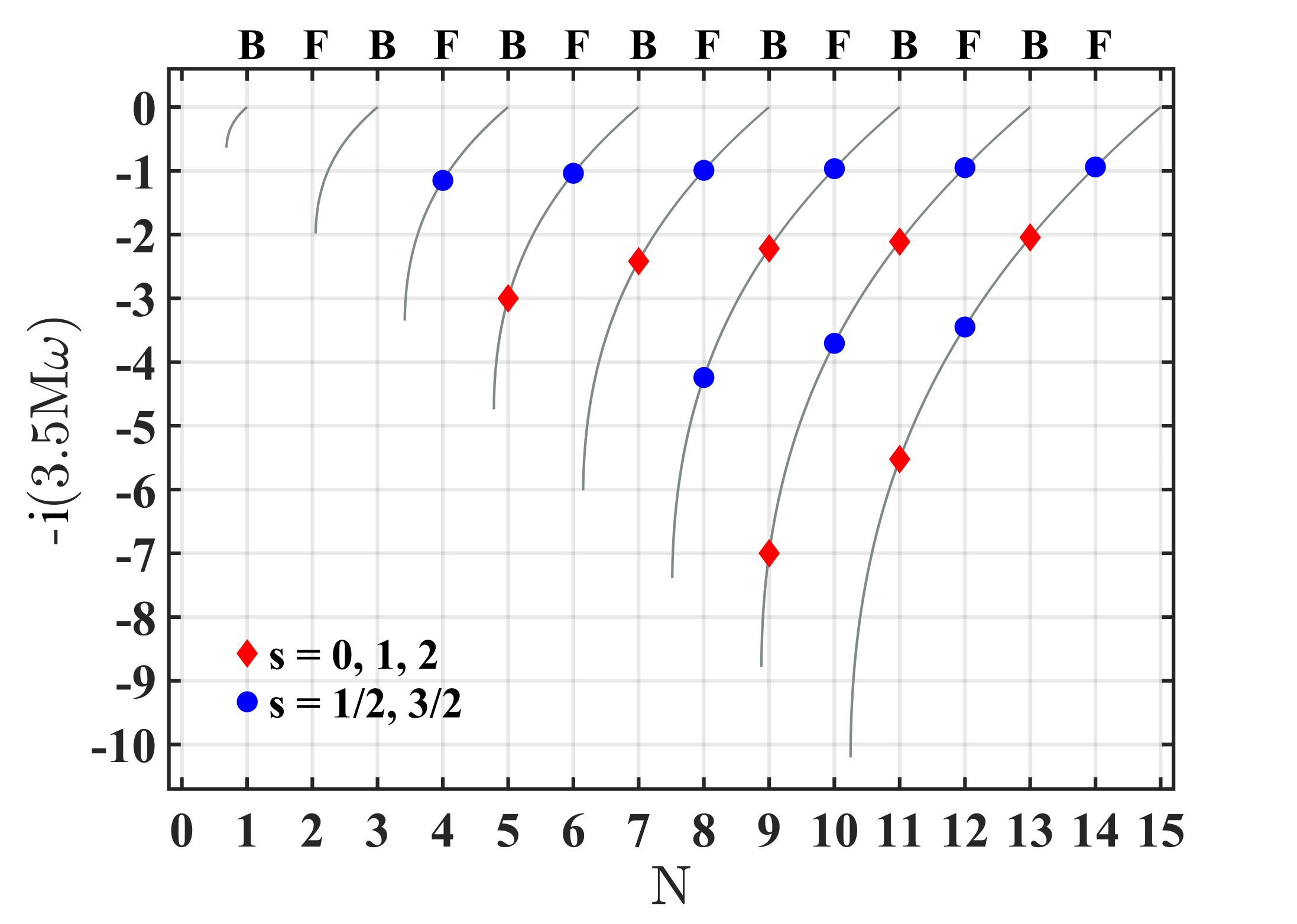}
\caption{\label{fig:2} Quasibound frequency spectrum with the smaller absolute value of $\omega$ in units of $1/(3.5M)$ for $N<2l+1<\sqrt{15/7}N$. $N$ is an odd integer for bosons (B) and an even integer for fermions (F). All frequencies on the same curve have the same quantum number $l$. From left to right, $l=0, 1, 2, 3, 4, 5, 6$, and $7$.}
\end{figure}


In this section, we have found solutions that match in the near-horizon and far-horizon regions. According to the uniqueness theorem for differential equations, the exact analytical solution must coincide with our derived solution in both the near-horizon and far-horizon regions. Therefore, the quasibound frequencies (53) obtained using the boundary conditions are correct, as it is guaranteed by the uniqueness of the solution to the differential equation.

\section{Discussion and conclusion}

It is well-known that the wave equations for different massless spin particles are rather different. Contrary to conventional wisdom, here we have found a single statement, Eq. (14), governing them, which shows that they have such a similar structure in Schwarzschild equivalent mediums. Equation (14) separates, and the solutions can be written as products of the time-dependent function, the angular function, and the radial function. The time-dependent function is $\mathrm{e}^{-\mathrm{i}\omega t}$, and the angular function is the spin-weighted spherical harmonics. The actual shape of the metric functions affects only the radial function. In particular, the function of $\Omega(r)$ in the solutions is a product factor of the radial function. Therefore, for a wave packet, the conformal factor affects the ``envelope", but does not affect the ``ripples".

We have investigated the radial equation in the near horizon and the far zones, and found that the radial equation can be transformed to a degenerate hypergeometric equation in any zone. Hence the solutions are governed by Kummer functions. Using the boundary condition (43), we have obtained the quasibound frequencies, which are given by Eq. (53). It is important to remember that $p=\pm s$. Thus Eq. (53) admits two kinds of particles: bosons, for which $N$ is an odd integer, and fermions, for which $N$ is an even integer. As one can see directly from Figs. 1 and 2, the quasibound frequencies of different bosons are identical at the same $N$ and $l$, and the same is true of different fermions, but a quasibound frequency for bosons can never equal a quasibound frequency for fermions. This fact provides a theoretical basis for simulation between particle-waves of same statistical properties.

Because fast development of metamaterials and transformation optics, one can realize beam shaping within dielectric slab samples with predesigned refractive index varying so as to create curved space environment for electromagnetic waves. As mentioned in the introduction, artificial optical materials have been proposed to study the various aspects of curved spacetimes. However, simulation of gravitational waves in curved spacetimes remains a challenge. Here our results provide a theoretical basis for electromagnetic waves are used to mimic gravitational waves. In particular, artificial optical materials have been realized to mimic Schwarzschild spacetime [41,55], therefore the transformation optics approach can be used to test our predictions.

\section*{Acknowledgments}
This work was supported by the National Natural Science Foundation of China under Grant No. 12175198.

\vspace*{20mm}
\appendix
\begin{center}
\LARGE\bf {Appendix}
\end{center}

\section{Spin Coefficients and Weyl Scalars}
\renewcommand{\theequation}{A\arabic{equation}}
\setcounter{equation}{0}

The 12 spin coefficients are defined by [56-58]
\begin{eqnarray}
&&\kappa=\bigtriangledown_{\nu}l_{\mu}m^{\mu}l^{\nu}, \quad \lambda=-\bigtriangledown_{\nu}n_{\mu}\bar{m}^{\mu}\bar{m}^{\nu},
\quad \sigma=\bigtriangledown_{\nu}l_{\mu}m^{\mu}m^{\nu}, \quad \nu=-\bigtriangledown_{\nu}n_{\mu}\bar{m}^{\mu}n^{\nu},\nonumber\\
&&\rho=\bigtriangledown_{\nu}l_{\mu}m^{\mu}\bar{m}^{\nu}, \quad \tau=\bigtriangledown_{\nu}l_{\mu}m^{\mu}n^{\nu},\quad \mu=-\bigtriangledown_{\nu}n_{\mu}\bar{m}^{\mu}m^{\nu}, \quad \pi=-\bigtriangledown_{\nu}n_{\mu}\bar{m}^{\mu}l^{\nu},\nonumber\\
&&\alpha=\frac{1}{2}(\bigtriangledown_{\nu}l_{\mu}n^{\mu}\bar{m}^{\nu}-\bigtriangledown_{\nu}m_{\mu}\bar{m}^{\mu}\bar{m}^{\nu}),\quad \beta=\frac{1}{2}(\bigtriangledown_{\nu}l_{\mu}n^{\mu}m^{\nu}-\bigtriangledown_{\nu}m_{\mu}\bar{m}^{\mu}m^{\nu}),\nonumber\\
&&\gamma=\frac{1}{2}(\bigtriangledown_{\nu}l_{\mu}n^{\mu}n^{\nu}-\bigtriangledown_{\nu}m_{\mu}\bar{m}^{\mu}n^{\nu}),\quad
\varepsilon=\frac{1}{2}(\bigtriangledown_{\nu}l_{\mu}n^{\mu}l^{\nu}-\bigtriangledown_{\nu}m_{\mu}\bar{m}^{\mu}l^{\nu}).
\end{eqnarray}\label{A1}
Here, $l_{\mu}$, $n_{\mu}$, $m_{\mu}$ and $\bar{m}_{\mu}$ are the Newman-Penrose null tetrad. The tetrad consists of two real null vectors, $l_{\mu}$ and $n_{\mu}$, and a pair of complex null vectors, $m_{\mu}$ and $\bar{m}_{\mu}$, which satisfies the orthonormal conditions, $l_{\mu}n^{\mu}=-m_{\mu}\bar{m}^{\mu}=1$, and $l_{\mu}l^{\mu}=n_{\mu}n^{\mu}=m_{\mu}m^{\mu}=\bar{m}_{\mu}\bar{m}^{\mu}=0$. The indexes are raised and lowered using global metric $g_{\mu\nu}$, which in terms of null vectors can be expressed as, $g_{\mu\nu}=2l_{(\mu}n_{\nu)}-2m_{(\mu}\bar{m}_{\nu)}$.

The Newman-Penrose tetrad for Schwarzschild equivalent mediums can be chosen as
\begin{eqnarray}
l_{\mu}&=&\frac{\Omega^{4}}{n}(\frac{1}{n}, -1, 0, 0),\nonumber\\
n_{\mu}&=&\frac{1}{2\Omega^{2}}(1, n, 0, 0),\nonumber\\
m_{\mu}&=&-\frac{\Omega r}{\sqrt{2}}(0, 0, 1, \mathrm{i}\sin\theta).
\label{A2}
\end{eqnarray}
Using the metric (5) and null tetrad (A2), the spin coefficients can be written as
\begin{eqnarray}
&&\kappa=\sigma=\nu=\lambda=\pi=\tau=0,\nonumber\\
&&\rho=-\frac{\Omega^{2}}{n}(\frac{\Omega'}{\Omega}+\frac{1}{r}),
\quad \mu=-\frac{n}{2\Omega^{4}}(\frac{\Omega'}{\Omega}+\frac{1}{r}),\nonumber\\
&&\varepsilon=\frac{\Omega^{2}}{n}(2\frac{\Omega'}{\Omega}-\frac{n'}{n}), \quad \gamma=-\frac{n}{2\Omega^{4}}\frac{\Omega'}{\Omega},\quad
\alpha=-\beta=-\frac{1}{2\sqrt{2}\Omega r}\cot\theta.
\label{A3}
\end{eqnarray}

The five Weyl scalars are defined by [56-58]
\begin{eqnarray}\label{A4}
&&\psi_{0}=-C_{\mu\nu\rho\sigma}l^{\mu}m^{\nu}l^{\rho}m^{\sigma},\quad
\psi_{1}=-C_{\mu\nu\rho\sigma}l^{\mu}n^{\nu}l^{\rho}m^{\sigma},\nonumber\\
&&\psi_{2}=-\frac{1}{2}C_{\mu\nu\rho\sigma}(l^{\mu}n^{\nu}l^{\rho}n^{\sigma}-l^{\mu}n^{\nu}m^{\rho}\bar{m}^{\sigma}),\nonumber\\
&&\psi_{3}=-C_{\mu\nu\rho\sigma}\bar{m}^{\mu}n^{\nu}l^{\rho}n^{\sigma},\quad
\psi_{4}=-C_{\mu\nu\rho\sigma}\bar{m}^{\mu}n^{\nu}\bar{m}^{\rho}n^{\sigma},
\end{eqnarray}
where $C_{\mu\nu\rho\sigma}$ is the Weyl tensor, which satisfies
\begin{equation}\label{A5}
 R_{\mu\nu\rho\sigma} =C_{\mu\nu\rho\sigma}+\frac{1}{2}(g_{\mu\rho}R_{\nu\sigma}-g_{\mu\sigma}R_{\nu\rho}-g_{\nu\rho}R_{\mu\sigma}\\
  +g_{\nu\sigma}R_{\mu\rho})+\frac{1}{6}(g_{\mu\sigma}g_{\nu\rho}-g_{\mu\rho}g_{\nu\sigma})R.
\end{equation}

\section{Spin Field Equations}
\renewcommand{\theequation}{B\arabic{equation}}
\setcounter{equation}{0}

The Weyl neutrino satisfies the wave equation [59]
\begin{equation}\label{B1}
\nabla_{AA'}P^{A}=0,
\end{equation}
where $P^{A}$ is the two-component spinor, $\nabla_{AA'}$ is the symbol for covariant spinor differentiation. Equation (B1) can be written in the Newman-Penrose formalism in the form
\begin{eqnarray}\label{B2}
&&(D+\varepsilon-\rho)P^{0}+(\bar{\delta}+\pi-\alpha)P^{1}=0,\nonumber\\
&&(\delta+\beta-\tau)P^{0}+(\Delta+\mu-\gamma)P^{1}=0.
\end{eqnarray}
In type-D spacetime, the equation can be decoupled into [58]
\begin{eqnarray}\label{B3}
&&[(D+\bar{\varepsilon}-\rho-\bar{\rho})(\Delta-\gamma+\mu)
-(\delta-\bar{\alpha}-\tau+\bar{\pi})(\bar{\delta}-\alpha+\pi)]\Phi_{+1/2}=0,\nonumber\\
&&[(\Delta-\bar{\gamma}+\mu+\bar{\mu})(D+\varepsilon-\rho)
-(\bar{\delta}+\bar{\beta}-\pi-\bar{\tau})(\delta+\beta-\tau)]\Phi_{-1/2}=0,
\end{eqnarray}
with $\Phi_{+1/2}=P^{1}$, and $\Phi_{-1/2}=-P^{0}$.

Similarly, the equations of  the electromagnetic ($s=1$), massless Rarita-Schwinger ($s=3/2$), and gravitational ($s=2$) fields on any type-D spacetime background can also be decoupled. For the source free case, they are given by [58,60]
\begin{eqnarray}\label{B4}
&&[(D-\varepsilon+\bar{\varepsilon}-2\rho-\bar{\rho})(\Delta-2\gamma+\mu)
-(\delta+\bar{\pi}-\bar{\alpha}-\beta-2\tau)(\bar{\delta}+\pi-2\alpha)]\Phi_{+1}=0,\nonumber\\
&&[(\Delta+\gamma-\bar{\gamma}+2\mu+\bar{\mu})(D+2\varepsilon-\rho)
-[\bar{\delta}-\bar{\tau}+\bar{\beta}+\alpha+2\pi)(\delta-\tau+2s\beta)\Phi_{-1}=0.
\end{eqnarray}
\begin{eqnarray}\label{B5}
&&[D-2\varepsilon+\bar{\varepsilon}-3\rho-\bar{\rho})(\Delta-3\gamma+\mu)
-(\delta+\bar{\pi}-\bar{\alpha}-2\beta-3\tau)(\bar{\delta}+\pi-3\alpha)
-\psi_2]\Phi_{+3/2}=0,\nonumber\\
&&[(\Delta+2\gamma-\bar{\gamma}+3\mu+\bar{\mu})(D+3\varepsilon-\rho)
-(\bar{\delta}-\bar{\tau}+\bar{\beta}+2\alpha+3\pi)(\delta-\tau+3\beta)
-\psi_2]\Phi_{-3/2}=0.\nonumber\\
\end{eqnarray}
\begin{eqnarray}\label{B6}
&&[(D-3\varepsilon+\bar{\varepsilon}-4\rho-\bar{\rho})(\Delta-4\gamma+\mu)
-(\delta+\bar{\pi}-\bar{\alpha}-3\beta-4\tau)(\bar{\delta}+\pi-4\alpha)
-3\psi_2]\Phi_{+2}=0,\nonumber\\
&&[(\Delta+3\gamma-\bar{\gamma}+4\mu+\bar{\mu})(D+4\varepsilon-\rho)
-[\bar{\delta}-\bar{\tau}+\bar{\beta}+3\alpha+4\pi](\delta-\tau+4\beta)
-3\psi_2]\Phi_{-2}=0.\nonumber\\
\end{eqnarray}

\end{document}